\documentclass[%
 reprint,
 amsmath,amssymb,
 aps,
]{revtex4-2}

\usepackage{graphicx}% Include figure files
\usepackage{dcolumn}% Align table columns on decimal point
\usepackage{bm}% bold math
\usepackage{xcolor}
\begin{document}

\preprint{APS/123-QED}

\title{ULTRAMAN:  A Proposed ultrasound detector for macroscopic dark matter}

\author{Oem Trivedi, Robert J. Scherrer, Alfredo Gurrola}
\affiliation{Department of Physics and Astronomy, Vanderbilt University, Nashville, TN, 37235, USA}
\email{Email: oem.trivedi@vanderbilt.edu \\ Email : robert.scherrer@vanderbilt.edu \\ Email: alfredo.gurrola@vanderbilt.edu }

\date{\today}% It is always \today, today,
             %  but any date may be explicitly specified

\begin{abstract}
%We propose ULTRAMAN (ULTRAsound Macro ANalyzer) a passive ultrasound array experiment to search for macroscopic dark matter (macros) through coherent acoustic disturbances generated as these compact objects traverse a controlled target medium. The framework links the macro stopping power and interaction cross section to the measurable pressure response, while the event rate probes the macro mass and abundance.A genuine candidate here would appear not as an isolated radio frequency spike, but as a correlated multichannel track consistent with a single highly supersonic trajectory through the target. Representative macro, nuclearite, axion quark nugget and primordial black hole benchmarks are shown to occupy distinct regions of the resulting acoustic observable space. Existing ultrasound technology offers a realistic route to an event level terrestrial search for macroscopic dark matter, with null observations translating into complementary constraints on its parameter space.
We propose ULTRAMAN (ULTRAsound Macro ANalyzer), a passive ultrasound array experiment designed to search for macroscopic dark matter (macros) through coherent acoustic disturbances generated as compact objects traverse a controlled target medium. Unlike existing dark matter detectors that primarily rely on scintillation, ionization, or other signatures of energy deposition, ULTRAMAN exploits the acoustic response of matter to reconstruct individual highly supersonic trajectories using synchronized multielement ultrasound arrays. The framework connects macro stopping power and interaction cross section to the resulting acoustic pressure, while the expected event rate probes the macro mass and abundance. A genuine candidate would appear not as an isolated radio frequency spike, but as a correlated multichannel track consistent with a single highly supersonic passage through the target. Representative macro, nuclearite, axion quark nugget, and primordial black hole benchmarks are shown to occupy distinct regions of the resulting acoustic observable space. Existing ultrasound technology offers a realistic route to an event-level terrestrial search for macroscopic dark matter, with null observations translating into complementary constraints on its parameter space.
\end{abstract}

\maketitle
Macroscopic dark matter represents a broad class of dark matter candidates in which the dark sector is organized into compact objects (macros) with masses far above the conventional particle dark matter scale. Such objects are commonly characterized by a mass $M_X$ and an effective interaction cross section $\sigma_X$ with ordinary matter. A particularly interesting possibility is that these objects are composite states, which means they include nuclear density objects, strange quark matter or nuclearites, axion or dark quark nuggets and other more general bound states of Standard Model or hidden sector particles \cite{mac9,mac10,mac11,mac12,mac13,mac14,mac15,mac16,mac17,macconstPicker:2026nuu}. Primordial black holes provide a conceptually different but related compact dark matter possibility in which the interaction with ordinary matter is primarily gravitational. These candidates are all very interesting because their macroscopic nature allows them to deposit energy along well localized trajectories through terrestrial matter, while significant regions of their mass and interaction parameter spaces remain difficult to access with conventional particle dark matter experiments . 

It is natural to consider the effects of macros on terrestrial matter, and this question has already motivated a number of unconventional approaches to dark matter detection. Sidhu, Scherrer and Starkman considered the consequences of macroscopic dark matter objects passing directly through the human body and showed that, for a sufficiently large interaction cross section, such passages could produce significant injury or even death \cite{sidhu2020death}. For a macro with velocity $v_X$ and geometric cross section $\sigma_X$ moving through a medium of density $\rho$, the energy deposited per unit path length can be written as
\begin{equation}
\label{Ex}
\frac{dE}{dx}=\sigma_X\rho v_X^2
\end{equation}
so that for a characteristic distance $L$ through the material one obtains
\begin{equation}
\label{Edep}
E_{\rm dep}=\sigma_X\rho v_X^2L
\end{equation}
This framework makes it clear that the terrestrial phenomenology of macros is not restricted to biological damage. The same energy deposition that could affect living tissue must first appear as a local mechanical, thermal or acoustic disturbance in the material through which the object passes. This immediately suggests that terrestrial matter itself may perhaps be used as a controlled target for detecting macroscopic dark matter.

This observation thus motivates us to consider
other possible detection mechanisms. Modern medical and laboratory diagnostics already employ highly sensitive physical probes of matter, and it is natural to ask whether these instruments themselves could respond to the passage of a macro even when the deposited energy is far below the threshold required to produce appreciable biological damage. In this sense, neither a human body nor any other living system needs to serve as the detector and one may instead use a controlled medium, such as water or a tissue-mimicking substance, as the target through which a macroscopic dark matter object could generate a transient signal that is subsequently measured by a precision detector. Such a construction also removes the ethical and practical limitations associated with using biological targets and allows the geometry, density and acoustic properties of the detection medium to be selected and calibrated experimentally. This is, however, not the only way in which macros are supposed to interact with terrestrial objects, since their physical effects can vary considerably depending on their composition and the nature of their interactions with ordinary matter. For example, sufficiently interacting macros traversing the Earth's atmosphere could generate luminous meteor-type events or leave behind distinctive, permanently modified or melted tracks. Other possibilities include antimatter macros depositing energy through matter-antimatter annihilation and primordial black holes generating seismic disturbances \cite{mac2,mac7,mac17,luo2012detectable}. In the present work, we focus on the geometric energy deposition framework of Sidhu, Scherrer, and Starkman, as motivated below.

Among standard medical imaging and probing methods, ultrasound is particularly well matched to the physics of macro dark matter energy deposition. In comparison, magnetic resonance imaging measures nuclear magnetization and spin relaxation in strong magnetic fields, and hence requires a mechanism through which a passing macro significantly perturbs the local magnetic environment or accumulated spin phase \cite{mri1,mri2}. Computed tomography instead measures integrated X ray attenuation and is not naturally optimized for detecting a brief mechanical disturbance caused by a compact object crossing the target \cite{ct1,ct2}. Single-photon emission computed tomography is sensitive to energetic photons produced by radioactive or other radiative processes, so its applicability is strongly dependent on the macroscopic dark matter model \cite{spect1,spect2}. Ultrasound, by contrast, directly measures pressure disturbances propagating through matter. Since a sufficiently fast macro deposits energy continuously along its trajectory, the resulting localized heating, displacement and pressure disturbance could provide a direct physical channel through which its passage could be recorded acoustically \cite{us1,us2,us3,us4}. This makes ultrasound particularly attractive as a comparatively model-independent probe of macroscopic objects that transfer appreciable energy to ordinary matter.

Consider the physical framework behind ultrasound techniques. The propagation of a small acoustic pressure perturbation $p(\mathbf{x},t)$ through approximately homogeneous tissue is described at leading order by the simple acoustic wave equation
\begin{equation}
\nabla^2p-\frac{1}{c_s^2}\frac{\partial^2p}{\partial t^2}=0.
\end{equation}
The quantity $c_s$ is the local speed of sound, which is approximately $1540\,{\rm m\,s^{-1}}$ in soft tissue. Medical ultrasound transducers generate and detect pressure oscillations at frequencies typically in the MHz regime. Spatial information is ordinarily obtained from the travel time of a transmitted pulse and its subsequent reflection from an acoustic interface.
For a reflector located at depth $d$, the return time is approximately $t_{\rm echo}={2d}/{c_s}$. This relation shows that conventional pulse-echo ultrasound determines the spatial location of reflecting interfaces by measuring the time delay between the transmission of an acoustic pulse and the reception of its reflected signal. The measured radio frequency signal may hence be represented schematically as a sum of reflected, band-limited responses together with instrumental and tissue noise
\begin{equation}
s_{\rm normal}(t)=\sum_j A_jh\left(t-\frac{2d_j}{c_s}\right)+n(t).
\end{equation}
Here $h(t)$ denotes the impulse response of the transducer, $A_j$ represents the effective reflectivity of a particular interface, and $n(t)$ contains electronic noise, speckle and other fluctuations. Normal radio frequency (RF) data hence contain a structured sequence of echoes whose timing is connected to previously transmitted acoustic pulses. A macro passage introduces a qualitatively different source term and the object itself gravitationally excites the medium, so the measured waveform becomes
\begin{equation}
s(t)=s_{\rm normal}(t)+s_{\rm MACRO}(t).
\end{equation}
The macro contribution need not be associated with any transmitted ultrasound pulse. It represents a spontaneous acoustic disturbance generated by the moving gravitational source and subsequently received by the ultrasound transducer. This absence of a corresponding transmit event is already one useful distinction from ordinary anatomical echoes.

With all that in mind, we now move towards discussing the architecture of our experimental proposal, which we call ULTRAMAN, for ULTRAsound Macro ANalyzer. The ULTRAMAN setup consists of a controlled target volume surrounded on several sides by multielement ultrasound arrays operated predominantly in a passive receiving mode. Rather than transmitting conventional imaging pulses continuously, the arrays would listen for spontaneous acoustic disturbances originating inside the target and record the raw RF response of each element with synchronized timing. Arrays placed above, below and along the lateral faces of the target would allow the same disturbance to be observed from several directions, making it possible to reconstruct whether the signals are consistent with a single compact object traversing the volume. The apparatus should be mechanically isolated from the laboratory environment and accompanied by auxiliary vibration, temperature and electromagnetic monitoring so that external mechanical disturbances and electronic artifacts can be vetoed. Active ultrasound measurements could still be performed intermittently for calibration and characterization of the target, while the dark matter search itself would rely on the long duration passive acquisition and the identification of coherent events across the full detector geometry. 

The choice of target material is equally important for our purposes because the experiment requires a medium that is acoustically well characterized, stable over long periods and sufficiently homogeneous that unusual transients are not dominated by internal imperfections. Degassed water is arguably the cleanest primary choice because its density and sound speed are precisely known, its attenuation is relatively low, it can be produced in large volumes at modest cost and removal of dissolved gas substantially suppresses bubble related acoustic backgrounds \cite{degwat}. If a solid tissue mimicking target is preferred, polyvinyl alcohol (PVA) cryogel is also attractive because it is mechanically durable and its acoustic properties can be tuned to resemble soft tissues \cite{pva} while remaining considerably more stable than ordinary gelatin. Agar or agarose provides another useful intermediate option, being comparatively easy to fabricate while offering good long term uniformity \cite{agar}. Gelatin remains valuable here for inexpensive proof of principle tests because it is simple to cast and acoustically tissue-like, but it is less suitable for a long duration search because dehydration, microbial growth and gradual mechanical changes can alter its response with time \cite{gelatin}. 

Hence, for the main experiment, one would favor degassed water for the cleanest and most scalable detector or PVA for a stable solid tissue-mimicker, while gelatin would be reserved for early calibration and prototype studies. A more ambitious future extension could also employ liquid xenon as the target medium, as its high density and well established use in large dark matter time projection chambers make it an interesting environment for enhanced macro energy deposition and complementary signal channels \cite{lz1akerib2020lux,xen1aprile2024xenonnt}. Its acoustic properties, including the speed of sound, have also been measured directly with pulse echo techniques, suggesting that a dedicated cryogenic acoustic implementation could in principle be calibrated in a controlled manner \cite{xensjaved2020thermodynamic}.
\begin{figure*}[t]
    \centering
    \includegraphics[width=0.8\linewidth]{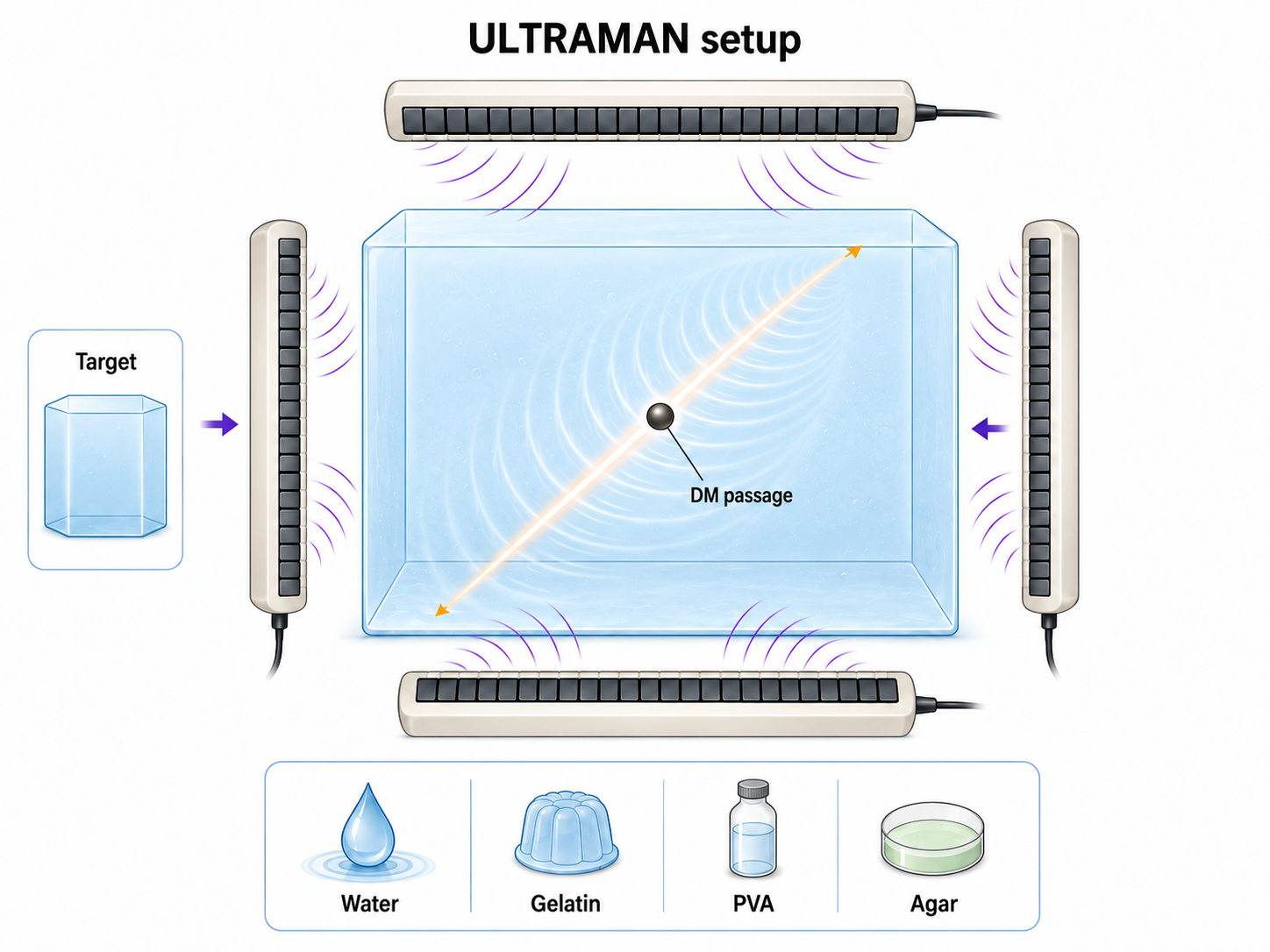}
    \caption{Schematic architecture of the ULTRAMAN passive ultrasound array experiment for detecting coherent acoustic disturbances from macroscopic dark matter passages through a controlled target medium.}
    \label{setup}
\end{figure*}
Fig.~\ref{setup} clarifies the ULTRAMAN setup, with a controlled target volume surrounded by multielement ultrasound arrays operated in passive mode. A traversing macroscopic dark matter object can deposit energy along a localized trajectory and generate an acoustic disturbance that propagates toward the surrounding receivers. The target medium can be chosen according to experimental requirements, with degassed water providing the cleanest primary option and gelatin, PVA and agar offering useful solid or tissue mimicking alternatives. The surrounding geometry is designed so that any genuine event can be observed redundantly from several directions rather than being inferred from a single local transient.

Having described the ULTRAMAN setup, we now develop the theoretical framework for the experiment. The most convenient model independent quantity is the stopping power $S_X$, defined as the energy deposited by a macroscopic dark matter object per unit path length through the target
\begin{equation}
S_X\equiv-\frac{dE_X}{dx}.
\end{equation}
For the geometric macro scenario considered in Ref.~\cite{sidhu2020death}, $S_X$ is given by Eq. (\ref{Ex}), and
for a path length $L$ through a homogeneous target, the corresponding deposited energy is given by Eq. (\ref{Edep}).

This immediately separates the two principal physical roles of the macro parameters. The interaction cross section controls how strongly an individual passage perturbs the target, while the mass $M_X$ primarily determines how frequently such objects cross the detector. Only a fraction of the deposited energy will appear in the acoustic modes ultimately recorded by a given ultrasound element. We denote this fraction by $\eta_X$ and collect the geometrical propagation, attenuation and receiver acceptance into a factor $\mathcal{G}_i$. The acoustic energy registered by element $i$ can then be written schematically as
\begin{equation}
E_{{\rm ac},i}=\eta_X\mathcal{G}_iE_{\rm dep}.
\end{equation}
For a pressure disturbance of characteristic duration $\Delta t_i$ received over an effective area $A_{{\rm eff},i}$, the acoustic energy is approximately
\begin{equation}
E_{{\rm ac},i}\simeq\frac{p_i^2}{\rho c_s}A_{{\rm eff},i}\Delta t_i,
\end{equation}
where $c_s$ is the sound speed of the target and combining these expressions gives us
\begin{equation}
p_i\simeq\left(\frac{\eta_X\mathcal{G}_iS_XL\rho c_s}{A_{{\rm eff},i}\Delta t_i}\right)^{1/2}
\end{equation}
and hence for geometric macros
\begin{equation}
p_i\simeq\rho v_X\left(\frac{\eta_X\mathcal{G}_i\sigma_Xc_sL}{A_{{\rm eff},i}\Delta t_i}\right)^{1/2},
\end{equation}
so that the characteristic scaling is
\begin{equation}
p_i\propto v_X\sqrt{\sigma_X}.
\end{equation}
The acoustic pressure generated by a traversing macro must be converted into a measurable electrical signal by the ultrasound transducers, whose response depends on their frequency-dependent sensitivity, bandwidth, directivity, and electronic noise. To account for these effects, we introduce a detector transfer function $H_i(f)$ for each array element, relating the incident acoustic pressure spectrum $P_i(f)$ to the recorded voltage spectrum through $V_i(f)=H_i(f)P_i(f)+N_i(f)$, where $N_i(f)$ represents the combined electronic and instrumental noise. The transfer function incorporates the transducer response and readout electronics, while acoustic propagation and attenuation within the target are already represented schematically by the geometrical factor $\mathcal{G}_i$ introduced above. Consequently, the pressure amplitudes estimated in Eqs. (9)–(11) should be interpreted as idealized incident acoustic signals rather than directly measured detector outputs. In practice, the observability of a macro passage depends on whether the resulting multichannel waveforms exceed the effective noise level within the relevant frequency band and satisfy the timing correlations expected for a common supersonic trajectory. For the present proof-of-concept proposal, we therefore characterize the detector sensitivity through a representative pressure threshold $p_{\mathrm{th}}$, understood as an effective threshold after accounting for the transducer response and instrumental noise. A more complete experimental implementation would determine $H_i(f)$, the noise spectral density, and the resulting detection efficiency $\epsilon_X(M_X,\sigma_X)$ through dedicated acoustic calibration measurements and simulated macro trajectories. These effects would modify the absolute detectable pressure and accessible macro parameter space, while preserving the underlying stopping-power scaling derived above under a fixed detector response.
The acoustic amplitude hence provides direct information about the stopping strength of the macro once the velocity and detector response are known. Other compact candidates can be incorporated simply by replacing $S_X$ with their appropriate stopping law. Primordial black holes would correspond to the gravitational limiting case here rather than geometric scattering. We also note that the macro abundance enters independently through the event rate. If macros constitute a fraction $f_X$ of the local dark matter density, their number density is
\begin{equation}
n_X=\frac{f_X\rho_{\rm DM}}{M_X}
\end{equation}
and the expected rate through an effective directional projected search area $A_{\rm search}$ is approximately
\begin{equation}
\Gamma_X=\frac{f_X\rho_{\rm DM}}{M_X}v_XA_{\rm search}\epsilon_X
\end{equation}
where $\epsilon_X$ denotes the complete event detection efficiency. The characteristic dependence $\Gamma_X\propto M_X^{-1}$ is especially useful, as the strength of an individual event is governed primarily by the stopping power, whereas its expected frequency carries information about the macro mass and abundance. A null result would hence constrain only the portion of $(M_X,\sigma_X)$ space for which both the acoustic response and the expected flux are sufficiently large.

Another essential observable is the coherent timing pattern across the ultrasound array. To calculate that, let the macro follow the trajectory
\begin{equation}
\mathbf{x}_X(t')=\mathbf{x}_0+\mathbf{v}_Xt'.
\end{equation}
A disturbance produced at time $t'$ reaches array element $i$ at position $\mathbf{x}_i$ at
\begin{equation}
t_i=t'+\frac{\left|\mathbf{x}_i-\mathbf{x}_X(t')\right|}{c_s}.
\end{equation}
Since the same trajectory generates the signals observed by all receiver elements, the measured arrival times must admit a common solution for $\mathbf{x}_0$, $\mathbf{v}_X$ and the passage time. This is the origin of the coherent track expected in channel and time space and provides a substantially stronger signature than an isolated RF spike. For Galactic macros one expects $v_X$ to be much larger than the acoustic propagation speed in the target, $v_X\gg c_s$, with the associated Mach geometry satisfying
\begin{equation}
\sin\theta_M=\frac{c_s}{v_X}
\end{equation}
The apparent sweep of the signal across the array should not itself be identified with $v_X$, since the pressure disturbance propagates through the target at $c_s$. Instead, the macro velocity can be obtained by fitting the complete set of retarded arrival times to a single supersonic trajectory.

Once the trajectory and velocity have been reconstructed, the measured pressure amplitude may also be inverted to estimate the stopping power
\begin{equation}
S_X\simeq\frac{p_i^2A_{{\rm eff},i}\Delta t_i}{\eta_X\mathcal{G}_i\rho c_sL}
\end{equation}
and, under the geometric macro hypothesis, the corresponding interaction cross section is
\begin{equation}
\sigma_X\simeq\frac{p_i^2A_{{\rm eff},i}\Delta t_i}{\eta_X\mathcal{G}_i\rho^2c_sLv_X^2}
\end{equation}
Thus the timing structure probes the trajectory and velocity, while the acoustic intensity probes the interaction strength. Multiple array elements would also provide redundant determinations of these quantities and hence would be an important internal consistency test. 

An additional consistency test can be obtained by placing two independently instrumented target volumes in series, for example one containing degassed water and another containing gelatin or PVA. For geometric stopping in two media,
\begin{equation}
S_{X,1}=\sigma_X\rho_1v_1^2
\end{equation}
and
\begin{equation}
S_{X,2}=\sigma_X\rho_2v_2^2
\end{equation}
this ends up giving us
\begin{equation}
\frac{S_{X,1}}{S_{X,2}}=\frac{\rho_1v_1^2}{\rho_2v_2^2}
\end{equation}
If the energy loss in the first target is small enough that $v_1\simeq v_2$, this simplifies to
\begin{equation}
\frac{S_{X,1}}{S_{X,2}}\simeq\frac{\rho_1}{\rho_2}
\end{equation}
\begin{figure*}[t]
    \centering
    \includegraphics[width=1\linewidth]{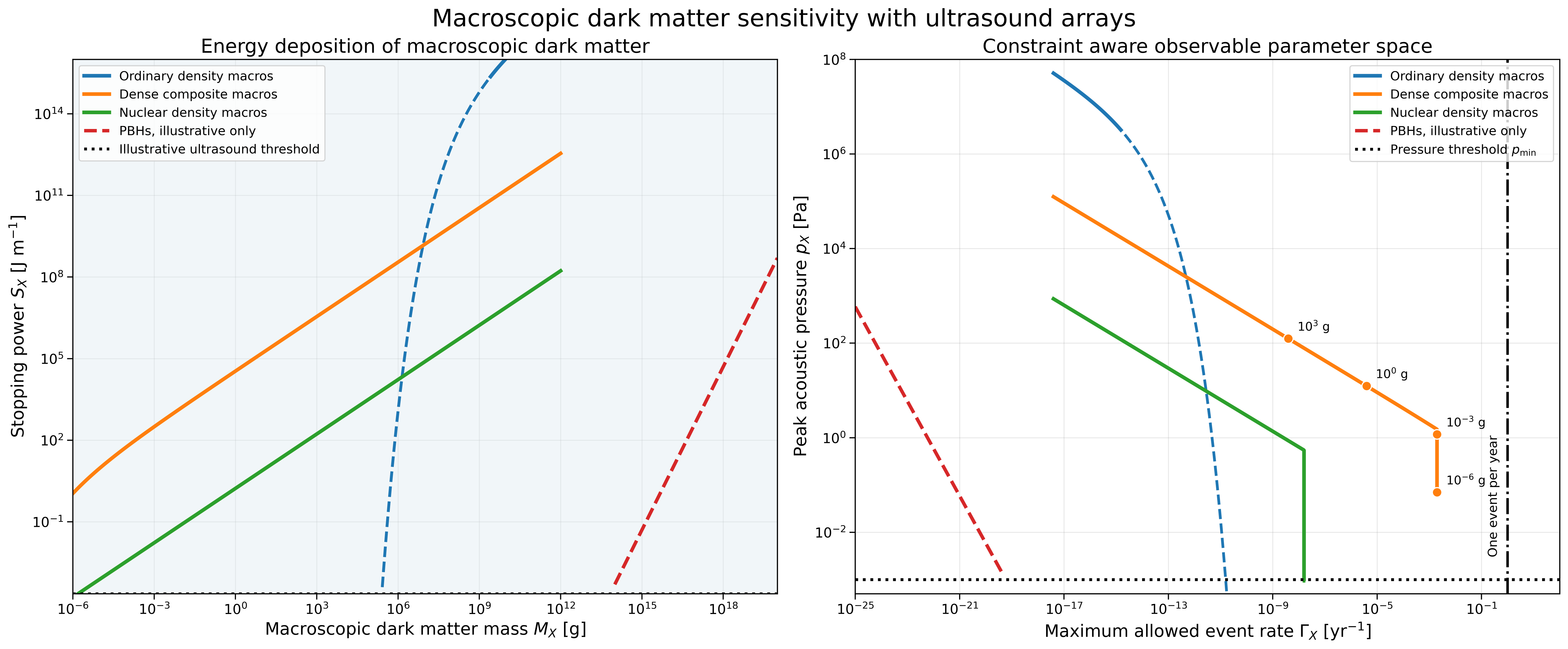}
    \caption{Illustrative stopping power and observable acoustic parameter space for representative macroscopic dark matter candidates in the proposed passive ultrasound array experiment. Note that the estimates here have been made taking into account the constraints on the maximum dark matter fractions allowed for various macros considered.}
    \label{ultarray}
\end{figure*}
The same reconstructed trajectory would also appear in both targets while the calibrated acoustic response changes consistently with the medium. Such a coincidence measurement would provide a useful way to discriminate a genuine traversing object from a local acoustic or instrumental background.

To illustrate the relevant scales as well, consider a $30\,{\rm cm}$ degassed water target with $\rho\simeq1000\,{\rm kg\,m^{-3}}$ and $c_s\simeq1480\,{\rm m\,s^{-1}}$ \cite{bilaniuk1993speed}, together with a representative Galactic velocity $v_X=250\,{\rm km\,s^{-1}}$ \cite{lewin1996review}. The macro crosses the target in approximately $1.2\,\mu{\rm s}$, while an acoustic wave takes roughly $200\,\mu{\rm s}$ to traverse the same distance, corresponding to a Mach number of about $170$. For a detector with $A_{\rm eff}=10^{-4}\,{\rm m^2}$, $\Delta t=1\,\mu{\rm s}$ and effective acoustic factor $\eta_X\mathcal{G}=10^{-18}$, a geometric macro with $\sigma_X=10^{-11}\,{\rm cm^2}$ gives $S_X\simeq0.063\,{\rm J\,m^{-1}}$ and an acoustic pressure scale of approximately $0.017\,{\rm Pa}$. Increasing the cross section to $\sigma_X=10^{-10}\,{\rm cm^2}$ gives $S_X\simeq0.63\,{\rm J\,m^{-1}}$ and $p\simeq0.053\,{\rm Pa}$, while $\sigma_X=10^{-9}\,{\rm cm^2}$ gives $S_X\simeq6.3\,{\rm J\,m^{-1}}$ and $p\simeq0.17\,{\rm Pa}$. These absolute pressure values depend on the presently uncalibrated acoustic conversion efficiency, whereas the scaling $p\propto\sqrt{\sigma_X}$ is considerably more robust. The corresponding experimental sensitivity will additionally depend on the frequency dependent detector response and noise characteristics discussed above. This means that the pressure threshold illustrated here should be regarded as a representative benchmark rather than an experimentally established detection limit. In the geometric macro examples shown below, we also account approximately for energy loss through a sea-level atmospheric column, and only emphasize trajectories which retain a sufficiently large velocity upon reaching the target.

The event rate exhibits the complementary mass dependence and taking $\rho_{\rm DM}\simeq5\times10^{-22}\,{\rm kg\,m^{-3}}$ \cite{baxter2021recommended}, $A_{\rm search}=1\,{\rm m^2}$ and $v_X=250\,{\rm km\,s^{-1}}$, macros with $M_X=10^{-6}\,{\rm g}$ would give approximately $3.9$ passages per year if they comprised all of the dark matter, whereas $M_X=10^{-4}\,{\rm g}$ gives about $3.9\times10^{-2}$ passages per year and $M_X=0.1\,{\rm g}$ gives only about $3.9\times10^{-5}$ passages per year. These values should be regarded as the corresponding $f_X=1$ reference rates, as in regions where existing searches already require the macro population to constitute only a fraction $f_X^{\rm max}<1$ of the dark matter, the maximum presently allowed rate is instead $\Gamma_X^{\rm max}=f_X^{\rm max}\Gamma_X(f_X=1)$. For details about the allowed fraction for macros, one can refer to \cite{picker2026macroscopic}. In Fig.~\ref{ultarray}, we therefore use representative existing abundance limits to rescale the event rate of the geometric macro benchmarks, so that the observable plane does not implicitly assume that every plotted candidate may comprise all of the dark matter. These estimates illustrate the basic experimental complementarity of our idea, as $\sigma_X$ controls whether a passage is acoustically visible, while $M_X$ controls whether such passages occur often enough to be observed.

Fig. \ref{ultarray} describes how representative macroscopic dark matter candidates map onto both their energy deposition in the target and the directly observable acoustic parameter space of the experiment. The left panel shows the stopping power trajectories of three representative constant density macro families, with atmospheric attenuation included and the more strongly drawn portions corresponding to trajectories that can reach the detector with a sufficiently large residual velocity. To take into account the attenuation, the incoming macro velocity is taken as \cite{mac2}
\begin{equation}
v_{\rm det}=v_X\exp\left(-\frac{\sigma_X\Sigma_{\rm atm}}{M_X}\right)
\end{equation}
where $\Sigma_{\rm atm}\simeq 1030\,{\rm g\,cm^{-2}}$ denotes the characteristic atmospheric column density, which gives us the stopping power at the detector as,
\begin{equation}
S_X=\rho_{\rm target}\sigma_X v_{\rm det}^{\,2}
\end{equation}
This allows us to capture the fact that macros with larger $\sigma_X/M_X$ lose a greater fraction of their incident velocity before reaching the target. Consequently, only trajectories retaining a sufficiently large residual velocity contribute to the experimentally accessible portions emphasized in the figure. The horizontal line denotes the illustrative acoustic sensitivity assumed in the present proof-of-concept estimate. The right panel maps the same candidates onto peak acoustic pressure and their maximum event rate consistent with the representative existing abundance bounds adopted here, making explicit the complementary roles of interaction strength and mass in determining detectability. In particular, lower mass macros can give us much larger event rates, although this advantage can be substantially reduced when existing searches already impose $f_X\ll1$.

Primordial black holes are retained in the figure only as an illustrative comparison of gravitational stopping and acoustic amplitude. They are not subjected to the same geometric scattering abundance prescription used for the other macro families, and their very small terrestrial flux makes them qualitatively different targets for a compact laboratory experiment. We would also like to emphasize that the benchmarks shown in Fig.~\ref{ultarray} are illustrative and do not necessarily represent entirely unconstrained dark matter scenarios, since existing constraints from Galactic positron annihilation, gamma ray observations and the cosmic X ray background already exclude primordial black holes around $10^{16}$-$10^{17}$ g from constituting all of the dark matter, with additional constraints extending to higher masses, while populations near $10^{18}$ g remain comparatively less restricted \cite{derocco2019constraining,berteaud2022strong,tan2024revisiting}. Similarly, nuclear density and other macroscopic dark matter candidates are subject to terrestrial track searches and cosmological constraints that exclude particular regions of their mass and interaction cross section parameter spaces too \cite{mac2,mac16}. The rates displayed here are intended as a conservative illustration of this effect rather than a replacement for a full recasting of every existing limit. The combined representation goes to show how a passive ultrasound search could both distinguish broad classes of macroscopic dark matter and identify the regions where sufficient signal strength and event rate overlap.
\begin{figure*}[t]
    \centering
    \includegraphics[width=0.9\linewidth]{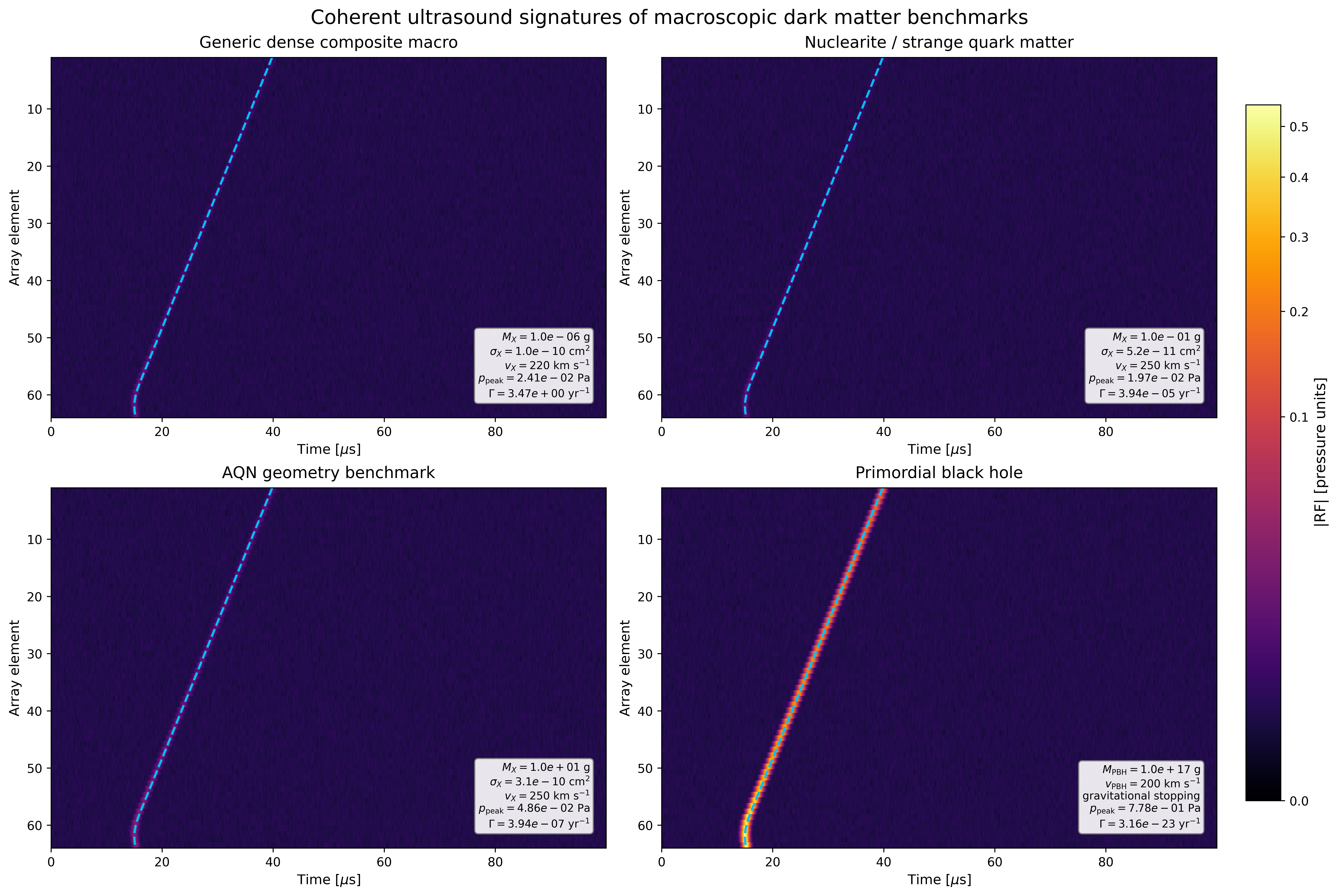}
    \caption{Illustrative coherent ultrasound array signatures for representative macroscopic dark matter candidates, showing their characteristic channel dependent arrival patterns and relative acoustic intensities.}
    \label{ultdm}
\end{figure*}

Perhaps even more striking is the signature illustrated in Fig. \ref{ultdm}, which shows the coherent channel and time response expected for several representative macroscopic dark matter benchmarks. In each case, the passage generates a continuous correlated ridge across the ultrasound elements, showing the common trajectory of a compact object through the target rather than an isolated transient in a single channel. The broadly similar track morphology follows from the fact that these candidates move with comparable Galactic velocities, while the relative RF intensity is controlled by their different stopping powers and interaction mechanisms. The nuclearite, dense composite and AQN benchmarks hence produce weaker but still coherent responses, whereas the PBH example gives a substantially larger acoustic amplitude under the adopted gravitational stopping prescription. The existence of such a coherent multichannel feature, together with its reconstructed velocity and intensity, constitutes the primary event level signature through which the experiment could identify and subsequently discriminate macroscopic dark matter candidates. 

Note that ULTRAMAN is realistic within present laboratory technology, since multielement ultrasound arrays, synchronized RF acquisition, passive acoustic imaging and tissue mimicking or water based substances are all already routinely available. The main experimental challenge would not be the basic hardware
but instead the ability to achieve sufficiently low acoustic and electronic backgrounds, stable long duration operation, and reliable calibration of the conversion between deposited energy and measured pressure. A first proof of principle experiment could be implemented using existing research ultrasound systems and a controlled water or PVA target, while a dedicated dark matter experiment would require larger target area and substantially longer exposure. In this sense, the proposal does not rely on speculative detector technology, but rather on adapting established ultrasound instrumentation to a new rare event search strategy. 

It is useful to place the present proposal in the context of other acoustic searches for ultraheavy dark matter. Cleaver et al.~\cite{cleaver2025listening} recently considered a conceptually related but experimentally very different approach in which macroscopic dark matter traversing seawater produces thermoacoustic pressure pulses through repeated nuclear scattering, with the signal subsequently detected by a sparse hydrophone array extending over a volume of order $100\,{\rm km^3}$. Their framework is naturally optimized for maximizing geometric exposure to extremely low flux candidates and consequently treats propagation over hundreds of meters to kilometers, including frequency dependent attenuation, dispersion and the detailed acoustic properties of seawater. Our proposal instead focuses on a compact and deliberately controlled target surrounded by dense multielement ultrasound arrays operated in passive mode. The central observable for us is not just the existence of an acoustic pressure pulse, but a densely sampled channel-dependent timing pattern that must be consistent with one common retarded trajectory and Galactic scale velocity. Related ideas have also been explored by Budker et.al. in the context of axion quark nuggets, where atmospheric, infrasonic, acoustic and seismic disturbances generated by macroscopic dark matter passages were proposed as possible signatures \cite{budker2020axion,budker2020axion}. These studies show the broader motivation for acoustic probes of macroscopic dark matter, while operating in detection regimes distinct from the controlled ultrasound architecture considered here.

This distinction is particularly important from the viewpoint of background control and experimental scalability. The underwater approach gains an enormous advantage in effective collection area, but necessarily operates in a natural environment containing sea state noise, biological and anthropogenic acoustic backgrounds and medium dependent attenuation, while requiring kilometer-scale instrumentation. The present form of ULTRAMAN sacrifices raw geometric exposure in exchange for substantially greater experimental control. In our case the target properties, receiver geometry, temperature, acoustic transfer function and electronic backgrounds can all be directly calibrated, while independent arrays and even two different media placed in series can provide redundant consistency tests for the same candidate trajectory. The apparatus is also modular and considerably easier to reproduce, allowing multiple installations to be operated at different laboratories without the infrastructure required by a large marine observatory. In this sense, the underwater proposal may be viewed primarily as an exposure maximizing search, whereas the present experiment is a control and information maximizing search aimed at establishing whether coherent acoustic tracking can provide a viable laboratory probe of macroscopic dark matter. Positive or null results from such controlled ULTRAMAN installations could consequently provide useful guidance for the interpretation, optimization and eventual motivation of much larger acoustic searches as well.

What would be required to claim a detection by ULTRAMAN? An individual RF spike would carry essentially no significance by itself, and even the observation of a coherent pattern of the form shown in Fig.~\ref{ultdm} in a single run would not be sufficient to establish macroscopic dark matter. A compelling case would require the same class of coherent multichannel event to be reconstructed across independent arrays, preferably across more than one target medium, with a common trajectory, a Galactic scale velocity, a consistent stopping power and no plausible acoustic, mechanical or electronic background explanation. Repeated events with mutually consistent properties would then begin to support the interpretation of a population of penetrating macroscopic objects, while the distribution of event rates and amplitudes could be used to test whether that population is compatible with a dark matter origin. Conversely, a null result over a calibrated exposure is also informative because it constrains only the region in which the experiment is demonstrably efficient, placing an upper bound
on $f_X$ as a function of $\sigma_X$ and $M_X$.
Thus, an absence of coherent events would exclude portions of the $(M_X,\sigma_X)$ space without ruling out macroscopic dark matter as a whole.

\textbf{Acknowledgements}: We gratefully acknowledge support from Vanderbilt
University and the U.S. National Science Foundation. The work of OT is supported in part by the Vanderbilt Discovery Doctoral Fellowship. The work of AG is supported in part by NSF Award PHY-2411502.
\bibliography{apssamp}% Produces the bibliography via BibTeX.

\bibliographystyle{apsrev4-2}
\end{document}